\documentclass[a4paper,12pt]{article}
\usepackage{amssymb}
\usepackage{amsmath}
\usepackage{epsf}
\usepackage{cite}

\makeatletter
\@addtoreset{equation}{section}
\makeatother

\def\preprint#1#2{\noindent\hbox{#1}\hfill\hbox{#2}\vskip 10pt}
\begin {document}
\begin{titlepage}

\preprint{ITP-UH-21/99}{December 1999}
\vfill

\begin{center}
{\Large\sc Magnetic properties of doped Heisenberg chains}
\vfill

{\sc Holger Frahm}$^{\dagger}$
        and
{\sc Nikita A.\ Slavnov}$^{\ddag}$
\vspace{1.0em}

$^{\dagger}${\sl
  Institut f\"ur Theoretische Physik, Universit\"at Hannover\\
  D-30167~Hannover, Germany}\\
$^{\ddag}${\sl
  Steklov Mathematical Institute, Moscow 117966, Russia}
\end{center}
\vfill

\begin{quote}
The magnetic susceptibility of systems from a class of integrable
models for doped spin-$S$ Heisenberg chains is calculated in the limit
of vanishing magnetic field.  For small concentrations $x_h$ of the
mobile spin-$(S-1/2)$ charge carriers we find an explicit expression
for the contribution of the gapless mode associated to the magnetic
degrees of freedom of these holes to the susceptibility which exhibits
a singularity for $x_h\to0$ for sufficiently large $S$.  We prove a
sum rule for the contributions of the two gapless magnetic modes in
the system to the susceptibility which holds for arbitrary hole
concentration.  This sum rule complements the one for the low
temperature specific heat which has been obtained previously.
\end{quote}

PACS-Nos.\ {71.10.Pm, 
	75.10.Jm, 
	75.10.Lp  
     }


\vspace*{\fill}
\setcounter{footnote}{0}
\end{titlepage}

%
\section{Introduction}
Essential insights into the properties of low-dimensional correlated
electron systems have been gained based on studies of integrable
models.  Exact results on the low energy spectrum of the
one-dimensional Hubbard and $t$--$J$ models have made possible their
identification as microscopic realizations of so-called
Tomonaga-Luttinger liquids allowing a complete classification of their
critical exponents \cite{FrKo90,FrKo91,KaYa91}.  Recently, the
$t$--$J$ model -- introduced to describe the effect of hole doping on
an $S=1/2$ Heisenberg antiferromagnet -- has been generalized to
larger values of $S$ \cite{DagX96,MHDa96} giving the framework for
detailed studies of quantum effects in the double exchange model.
Fine tuning of the coupling constants in these models describing
spin-$S$ chains doped with \emph{mobile} spin-$S'=(S-1/2)$ carriers
(called `holes' below) an exactly solvable model has been constructed
\cite{FrPT98,Frahm99} which interpolates smoothly between the spin-$S$
(spin-$S'$) Takhtajan-Babujian chain \cite{Takh82,Babu82,Babu83}.

An important open question is the proper identification of effective
field theories describing the low energy/low temperature behaviour of
these systems \cite{FrPT98,Frahm99,HoCo99,Saleur99}.  In the limit of
low hole concentration the low temperature specific heat indicates
that the two gapless magnetic modes found for vanishing magnetic field
can be described in terms of a level-$2S$ $SU(2)$ Wess-Zumino-Witten
(WZW) model (which is the effective field theory of the undoped
Takhtajan-Babujian spin-$S$ chain \cite{Takh82,Babu82,Babu83}) and the
minimal model ${\cal M}_{2S+1}$, respectively \cite{FrPT98,Frahm99}.
Upon variation of the doping a continous transition to a product
$SU(2)_{2S-1} \otimes SU(2)_1$ satisfying a sum rule in the
coefficients of the low temperature specific heat is observed
\cite{Frahm99}.  For $S=1$ a field theoretical description of these
degrees of freedom has been proposed based on four Majorana fermions
\cite{FrPT98} (see also \cite{HoCo99,Saleur99}).  However, to
determine the coupling constants present in this continuum theory
additional insight into the physical properties of the massless
excitations, in particular their response to an external magnetic
field, is necessary.

In this paper we analyze the Bethe Ansatz equations to compute the
zero temperature magnetic susceptibility of the integrable doped
Heisenberg chains in the limit of vanishing magnetic field $H\to0$.
Below we briefly review the thermodynamic Bethe Ansatz for the model
to provide the foundation for the subsequent analysis.  In
Sect.~\ref{sec:H0} we concentrate on the limit of small magnetic
fields and derive a particular sum rule for the contributions $\chi_1$
and $\chi_{2S}$ of the two gapless magnetic excitations to the
susceptibility, similar to the one for the specific heat mentioned
above.  In Sect.~\ref{sec:RHP} we formulate a matrix Riemann-Hilbert
problem (RHP) whose solution determines $\chi_1$ and $\chi_{2S}$
individually.  It is shown that the symmetries of this RHP immediately
imply the sum rule mentioned above.
In Sect.~\ref{sec:smallx} we present the main result of this paper
which is the explicit calculation of $\chi_1$ and $\chi_{2S}$ in the
limit of small hole concentration.

The Hamiltonians considered in this paper are of the form
\begin{equation}
   {\cal H}^{(S)} =\sum_{n=1}^L \left\{
	{\cal T}^{(S)}_{n,n+1}
      + {\cal X}^{(S)}_{n,n+1}\right\}\,.
\label{hamil1}
\end{equation}
Here ${\cal T}^{(S)}_{ij}$ and ${\cal X}^{(S)}_{ij}$ describe the
hopping of the holes and the (antiferromagnetic) exchange between
sites $i$ and $j$ of the lattice, respectively.  Similar Hamiltonians
arise as effective spin models obtained from the double exchange (DE)
model \cite{Zener51,AnHa55} in the limit of strong ferromagnetic
Hund's rule coupling between the spins of the itinerant electrons and
the local moments with spin-$S'$ \cite{DagX96,MHDa96}.
In the basis of the relevant spin multiplets the kinetic part of the
Hamiltonian reads
\begin{equation}
  {\cal T}^{(S)}_{ij}
   = - {\cal P}_{ij} Q_S(\mathbf{S}_i\cdot\mathbf{S}_j)\,
\label{Heleff}
\end{equation}
where $\mathbf{S}_{i}$ is a spin-$S$ ($S'$) operator describing the
particle (hole) on site $i$.  The operator ${\cal P}_{ij}$ permutes
the states on sites $i$ and $j$ thereby allowing the holes to move and
$Q_S(x)$ is a polynomial of degree $2S-1$ leading to a hopping
amplitude depending on the total spin $S_T= \frac{1}{2}, \frac{3}{2},
\ldots 2S-\frac{1}{2}$ on sites $i$ and $j$.  In the DE model these
hopping amplitudes favour the formation of a ferromagnetically ordered
state, this feature is not shared by the integrable chains
\cite{FrPT98,Frahm99}.  In terms of a microscopic model the electronic
hopping has to be dressed by bilinears in the localized magnetic
moments, similar to bond-charge type interactions considered in
generalizations of the Hubbard model.

As a consequence of $SU(2)$ invariance the exchange operators ${\cal
X}^{(S)}$ can also be written as a polynomial in $\mathbf{S}_i\cdot
\mathbf{S}_j$, the precise form of these polynomials depends on the
local configuration, i.e.\ particle-particle, hole-hole and
particle-hole exchange are different.  As mentioned above, the pure
limits are the integrable spin-$S$ and $S'$ Takhtajan-Babujian chains
determining the two former processes.

%
%
\section{Bethe Ansatz for the doped spin-$S$ chain}
\label{sec:TBA}
Starting from the ferromagnetically polarized eigenstate of the
undoped chain we consider excitations obtained by adding $N_h$ holes
and -- in addition -- lowering $N_\downarrow$ spins.  The spectrum of
these states can be studied using the algebraic Bethe Ansatz.  They
are parameterized in terms of $N_h$ real numbers $\nu_\alpha$ and
$N_h+N_\downarrow$ complex numbers $\lambda_j$ solving the Bethe
Ansatz equations (BAE) \cite{FrPT98,Frahm99}
\begin{eqnarray}
   \left( \frac{\lambda_j+iS}{\lambda_j-iS} \right)^L &=&
       \prod_{k\ne j}^{N_h+N_\downarrow}
	{{\lambda_j-\lambda_k+i}\over{\lambda_j-\lambda_k-i}}\,
       \prod_{\alpha=1}^{N_h}
	{{\lambda_j-\nu_\alpha-{i\over2}} \over
	 {\lambda_j-\nu_\alpha+{i\over2}} }\ ,
\nonumber\\
	&& \qquad j=1,\ldots,N_h+N_\downarrow
\label{bae3}\\
   1 &=& \prod_{k=1}^{N_h+N_\downarrow}
	{{\nu_\alpha - \lambda_k +{i\over2}} \over
	 {\nu_\alpha - \lambda_k -{i\over2}}}\ ,
   \quad \alpha=1,\ldots,N_h\ .
\nonumber
\end{eqnarray}
In the thermodynamic limit $L\to\infty$ general solutions of these
equations are known to consist of real hole rapidities $\nu_\alpha$
and complex $n$-\emph{strings} of spin-rapidities $\lambda_j^{n,k} =
\lambda_j^{(n)} + {i\over2}\left(n+1-2k\right)$, $k=1,\ldots,n$ with
real centers $\lambda_j^{(n)}$.  Considering solutions of (\ref{bae3})
built from $N_h$ hole rapidities and $M_n$ $\lambda$-strings of length
$n$ we can rewrite the BAE in terms of the real variables $\nu_\alpha$
and $\lambda_j^{(n)}$.  Taking the logarithm of (\ref{bae3}) we obtain
\begin{eqnarray}
   L\theta_{n,2S}\left({\lambda_j^{(n)}}\right) &=&
	{2\pi}\ J_j^{(n)}
	+\sum_{m=1}^\infty \sum_{k=1}^{M_{m}}
	 \Xi_{nm}\left(\lambda_j^{(n)}-\lambda_k^{(m)}\right)
	-\sum_{\alpha=1}^{N_h}
	 \theta_n\left({\lambda_j^{(n)}-\nu_\alpha}\right)
\nonumber\\
  0 &=& {2\pi}\ I_\alpha
     - \sum_{n=1}^\infty \sum_{j=1}^{M_n}
	\theta_n\left({\nu_\alpha-\lambda_j^{(n)}}\right)
\label{zz}
\end{eqnarray}
where $\theta_n(x) = 2\arctan(2x/n)$ and
\begin{eqnarray}
  \theta_{nm}(x) &=& \theta_{m+n-1}\left(x\right)
	+ \theta_{m+n-3}\left(x\right)+\ldots
	+ \theta_{|m-n|+1}\left(x\right)\ ,
\nonumber\\
  \Xi_{nm}(x) &=& \theta_{n+m}\left({x}\right)
 	+2\theta_{n+m-2}\left({x}\right)+\ldots
	+2\theta_{|n-m|+2}\left({x}\right)
	+\left(1-\delta_{nm}\right)\theta_{|n-m|}\left({x}\right)\
 	.
\end{eqnarray}
{}From Eqs.~(\ref{zz}) the allowed values of the quantum numbers
$I_\alpha$ and $J_j^{(n)}$ are found to be ($t_{nm} =
\min(n,m)-\delta_{nm}/2$)
\begin{equation}
  \left| I_\alpha \right| \le {1\over2} \sum_{m=1}^\infty M_n\ ,
\quad
  \left| J_j^{(n)}\right| \le {1\over2} L \min(n,2S)
	- \sum_{m=1}^\infty t_{nm} M_m
        + {1\over2} \left(N_h - 1\right)\ .
\label{ranges}
\end{equation}
A given choice of these numbers the solution of (\ref{zz}) uniquely
determines a particular eigenstate of the system with energy
\begin{equation}
  E = \sum_{n=1}^\infty \sum_{j=1}^{M_n}
	   \left\{\epsilon_n^{(0)}\left(\lambda_j^{(n)}\right)+nH\right\}
      - \left(\mu +{1\over2} H\right) N_h
      - LSH
\label{EEE}
\end{equation}
and magnetization
\begin{equation}
  S^z_{\rm tot} = LS  - \sum_{n=1}^\infty n M_n+ {1\over2} N_h \ .
\label{SZZ}
\end{equation}
In Eq.~(\ref{EEE}) $H$ is the external magnetic field, $\mu$ is the
chemical potential for the holes and the ``bare energies'' of the
$\lambda$-strings are $\epsilon_n^{(0)}(x)= -2\pi\left(A_{n,2S}\ast
s\right)(x)$ with $s(x) = 1/(2\cosh\pi x)$ and
\begin{equation}
  A_{nm}(x) = {1\over2\pi} \Xi_{nm}'(x) + \delta_{nm}\,\delta(x)\
\end{equation}
($\left(f\ast g\right)(x)$ denotes a convolution).

In the thermodynamic limit $L\to\infty$ with $N_h/L$ and $M_n/L$ held
fixed one can introduce densities $\rho(x)$ for the hole rapidities
and $\sigma_n(x)$ for the $\lambda$-strings of length $n$.  The BAE
(\ref{zz}) turn into  linear integral equations for these functions
\begin{eqnarray}
  &&\tilde\sigma_n(x) =
	\left(A_{n,2S}\ast s\right)(x)
	-\sum_m \left(A_{nm}\ast\sigma_m\right)(x)
	+\left(a_n\ast\rho\right)(x)\ ,
\nonumber\\
  &&\rho(x) + \tilde\rho(x) = \sum_n \left(a_n\ast\sigma_n\right)(x)\ .
\label{intd1}
\end{eqnarray}
Here, $2\pi a_n(x) = \theta_n'(x) = 4n/(4x^2 + n^2)$, and
$\tilde\rho(x)$, $\tilde\sigma_n(x)$ are the densities associated with
the distribution of vacancies of quantum numbers $I_\alpha$ and
$J_j^{(n)}$ in the intervals (\ref{ranges}), respectively.
Similarly, the energy (\ref{EEE}) and magnetization (\ref{SZZ})
expressed in terms of these densities read
\begin{equation}
  E/L = \sum_{n=1}^{\infty}
	\int{\rm d}x\,\epsilon_n^{(0)}(x)\sigma_n(x)
	 - \mu\,\int{\rm d}x\,\rho(x)
	-H\ S^z_{\rm tot}\ ,
\label{EET}
\end{equation}
and\footnote{The second expression is obtained by integrating
Eq.~(\ref{intd1}) over all $x$}
\begin{equation}
  {1\over L}\ S^z_{\rm tot} = S -
    \sum_{n=1}^{\infty} n\int{\rm d}x\,\sigma_n(x) +
    {1\over2}\int{\rm d}x\,\rho(x)\
  = {1\over2}\lim_{n\to\infty} \int{\rm d}x\,\tilde\sigma_n(x)\ .
\label{SZT}
\end{equation}

Analysis of these equations shows that the ground state configuration
of the system at temperature $T=0$ consists of up to three condensates
formed by the hole rapidities and $\lambda$-strings of length $1$ and
$2S$ only.  These ``Fermi seas'' are characterized by their end points
$Q$ for $\rho(x)$ and $\Lambda_j$ for $\sigma_j(x)$, $j=1,2S$.  In the
following we drop the distinction of particle and vacancy densities by
identifying $\rho(x)$ with $\rho(x)+\tilde\rho(x)$ and similar for the
densities $\sigma_n(x)$ of the $\lambda$-strings.  This is possible
since
\begin{equation}
  \rho(x)+\tilde\rho(x)\equiv\left\{
    \begin{array}{ll}
     \rho(x)&{\rm for~}|x|<Q\\
     \tilde\rho(x)&{\rm for~}|x|>Q
    \end{array}\right.\ .
\end{equation}
Using $M_j=0$ for $j\ne1,2S$ in Eqs.~(\ref{ranges}) and (\ref{SZZ})
one can express the magnetization in terms of $\sigma_{2S}$ only
\begin{equation}
 {1\over L}\, S^z_{\rm tot} = \int_{\Lambda_{2S}}^\infty {\rm d}x\
 \sigma_{2S}(x)\ .
\label{SZT0}
\end{equation}

%
\section{Thermodynamic properties in a small magnetic field}
\label{sec:H0}
For vanishing magnetic field one can show that $\Lambda_j\to\infty$.
Hence, in the limit of a small magnetic field $H$ which we are
interested in below one has $\Lambda_j\gg Q$ allowing to eliminate the
$\sigma_j$ from the integral equation for the density $\rho$ of hole
rapidities \cite{Frahm99}.  This results in
\begin{equation}
  \rho(x) - \int_{-Q}^Q {\rm d}y\,R(x - y)\rho(y)
  = \left(a_{2S}\ast s\right)(x)\, .
\label{rho0}
\end{equation}
The kernel of this integral equation is $R(x) = a_2\ast\left(1+
a_2\right)^{-1}$ and the concentration of the holes is $x_h=N_h/L
=\int_{-Q}^Q {\rm d}x\,\rho(x)$.  In this regime the boundaries of
integration or ``Fermi points'' $\pm Q$ are a function of the hole
chemical potential $\mu$ alone: they are determined through the
condition $\kappa_0(\pm Q)=0$ for the dressed energy $\kappa_(x)$ of the
excitations associated with the hole rapidities which in turn is given
as solution of a linear integral equation similar to (\ref{rho0})
\begin{equation}
   \kappa_0(x) -\int_{-Q}^{Q}{\rm d}y\,R(x-y)\kappa_0(y)=
	-\left\{ 2\pi a_{2S}\ast s(x) + \mu\right\}\ .
\label{kappa0}
\end{equation}
Excitations with charge rapidities near $\pm Q$ are massless.  The
effective low energy theory for the charge mode has been identified as
that of a free boson.  The velocity of the charge mode can be obtained
from the dispersion (\ref{kappa0})
\begin{equation}
  v = \frac{1}{2\pi\rho(Q)}
	\left.\frac{\partial\kappa_0}{\partial x}\right|_{x = Q}\ .
\label{v}
\end{equation}

The integral equations for the relevant densities of $\lambda$-strings
$\sigma_1$ and $\sigma_{2S}$ can be rewritten as
\begin{eqnarray}
  \sigma_1(x) &=& \int_{-Q}^Q{\rm d}y\ s(x-y) \rho(y)
	+ \int_{|y|>\Lambda_1}{\rm d}y\ K_{11}(x-y)\sigma_1(y)
	+ \int_{|y|>\Lambda_{2S}}{\rm d}y\
                      K_{12}(x-y)\sigma_{2S}(y)\ ,
\nonumber\\
\label{intds}\\
  \sigma_{2S}(x)  &=& s(x)
	+ \int_{|y|>\Lambda_1}{\rm d}y\
		    K_{21}(x-y)\sigma_1(y)
	+ \int_{|y|>\Lambda_{2S}}{\rm d}y\
	            K_{22}(x-y)\sigma_{2S}(y)\ .
\nonumber
\end{eqnarray}
The kernels of the integral operators are easiest given in terms of
their Fourier transforms
\begin{equation}
  \hat\mathbb{K}^{(S)}(\omega)
 = {1\over2\cosh{1\over2}\omega\,\sinh(S-{1\over2}){\omega}}
   \left( \begin{array}{cc}
     \sinh(S-1)\omega & \sinh{1\over2}\omega \\
     \sinh{1\over2}\omega & \sinh(S-1)\omega +
        {\rm e}^{-{|\omega|\over2}}\sinh(S-{1\over2})\omega
	  \end{array}\right)\ .
\label{kernS}
\end{equation}
Note that for any given positive $x$ the contributions from the
intervals $y<-\Lambda_j$ in Eqs.~(\ref{intds}) can be neglected in the
limit $H\to0$.  After replacing $\int_{|y|>\Lambda_j}$ by
$\int_{\Lambda_j}^\infty$ these equations are a system of Wiener-Hopf
(WH) integral equations.  Furthermore, the replacement of the driving
terms by the large-$x$ asymptotics will not change the behaviour of
the solutions of these equations near the Fermi points
$x\approx\Lambda_j\gg1$ which contains all the relevant information
for the low energy properties of the system.  Hence we arrive at the
following set of equations for the densities of the $\lambda$-strings
valid for $x\gg1$
\begin{eqnarray}
  \sigma_1(x) &=& B\,e^{-\pi |x|}
	+ \int_{\Lambda_1}^\infty{\rm d}y\
	              K_{11}(x-y)\sigma_1(y)
	+ \int_{\Lambda_{2S}}^\infty{\rm d}y\
                      K_{12}(x-y)\sigma_{2S}(y)\ ,
\nonumber\\
\label{WHdens}\\
  \sigma_{2S}(x)  &=& e^{-\pi |x|}
	+ \int_{\Lambda_1}^\infty{\rm d}y\
		    K_{21}(x-y)\sigma_1(y)
	+ \int_{\Lambda_{2S}}^\infty{\rm d}y\
	            K_{22}(x-y)\sigma_{2S}(y)\ .
\nonumber
\end{eqnarray}
with $B=\int_{-Q}^Q{\rm d}x\ \exp(\pi x) \rho(x)$.
In the same spirit we obtain a system of WH equations for the dressed
energies of the magnetic excitations
\begin{eqnarray}
  \epsilon_1(x) &=& -2\pi A\,e^{-\pi |x|}
	+ \int_{\Lambda_1}^\infty{\rm d}y\
	              K_{11}(x-y)\epsilon_1(y)
	+ \int_{\Lambda_{2S}}^\infty{\rm d}y\
                      K_{12}(x-y)\epsilon_{2S}(y)\ ,
\nonumber\\
\label{WHener}\\
  \epsilon_{2S}(x)  &=& {1\over2}H-2\pi\,e^{-\pi |x|}
	+ \int_{\Lambda_1}^\infty{\rm d}y\
		    K_{21}(x-y)\epsilon_1(y)
	+ \int_{\Lambda_{2S}}^\infty{\rm d}y\
	            K_{22}(x-y)\epsilon_{2S}(y)\ .
\nonumber
\end{eqnarray}
with $2\pi A=-\int_{-Q}^Q{\rm d}y\ \exp(\pi x) \kappa_0(y)$.  The
solution of (\ref{WHener}) determines the values of $\Lambda_j$ as a
function of the external magnetic field $H$ by the condition
$\epsilon_j(\Lambda_j) = 0$.

Both magnetic modes $\epsilon_{1}(x)$ and $\epsilon_{2S}(x)$ allow for
massless excitations near these Fermi points, in the limit $H\to0$
their velocities are
\begin{equation}
  v_{2S} = \lim_{x\to\infty}
	{\epsilon_{2S}'(x)\over2\pi\sigma_{2S}(x)} \equiv \pi\ ,
\quad
  v_{1} = \lim_{x\to\infty}
	{\epsilon_{1}'(x)\over2\pi\sigma_1(x)}
	= \pi\, \frac{A}{B}\ .
\label{veloS}
\end{equation}
Their critical properties have been studied using the thermodynamic
Bethe Ansatz.  In the limiting cases of the undoped (completely doped)
model where the model reduces to the integrable spin-$S$
(spin-($S-1/2$)) Takhtajan-Babujian spin chains, respectively,
$\epsilon_{2S}(x)$ reduces to the massless spinon mode of these models
which can be described by a $SU(2)$ level-$2S$ (level-($2S-1$))
WZW model.  For arbitrary doping the contribution of the magnetic
modes to the low temperature specific heat has been found to be
\cite{FrPT98,Frahm99}
\begin{equation}
   C \simeq C_{2S} + C_1 \equiv {\pi T\over3}
    \left( {c_{2S}\over v_{2S}} + {c_1\over v_1}\right)
\label{spech}
\end{equation}
where the parameters $c_j$ satisfy the ``sum rule''
\begin{equation}
  c_{2S}+c_1 \equiv 2\,{4S-1\over 2S+1}\
\label{sumC}
\end{equation}
independent of the doping, i.e.\ the values of the parameters $A$, $B$
in Eqs.~(\ref{WHdens}) and (\ref{WHener}).  In the limiting cases
mentioned above $c_{2S}$ becomes the conformal central charge of the
$SU(2)_{2S}$ and the $SU(2)_{2S-1}$ WZW models, respectively.  From
the other parameter, $c_1$, the low energy theory for $\epsilon_1$ is
identified as the minimal unitary model ${\cal M}_{2S+1}$ in the limit
of vanishing hole concentration ($x_h\to0$) and a free boson with
$c_1=1$ and compactification radius $R=1/\sqrt{2}$ as $x_h\to1$
\cite{Frahm99}.  The latter is equivalent to the level-$1$ $SU(2)$ WZW
model.

The magnetic field dependent part of ground state energy of the system
for small $H$ expressed in terms of the solutions of (\ref{WHener}) is
\begin{equation}
  e_s \equiv {1\over L} E = -2\pi \left(
   {1\over v_{2S}}\int_{\Lambda_{2S}}^\infty{\rm d}x\
	e^{-\pi x}\epsilon_{2S}(x)
  +{1\over v_1}\int_{\Lambda_1}^\infty{\rm d}x\
	A\,e^{-\pi x}\epsilon_1(x) \right)
  = -{1\over2} \chi\, H^2 .
\label{Emag}
\end{equation}
This implies, that the zero field magnetic susceptibility $\chi$ will
have contributions from both massless magnetic modes -- very similar
to the feature found for the specific heat (\ref{spech})
\begin{equation}
  \chi = \chi_{2S} + \chi_1
\label{susc0}
\end{equation}
A sum rule for the contributions $\chi_j$ can be obtained by
considering formally the case $A=B$ which implies $v_1=v_{2S}=\pi$:
then the solutions of the integral equations (\ref{WHdens}) and
(\ref{WHener}) are related through
\begin{equation}
  \sigma_j(x) = {1\over2\pi^2}\,
	{\partial\epsilon_j(x)\over\partial x}\ .
\end{equation}
Using this relation with the asymptotic behaviour $\lim_{x\to\infty}
\epsilon_{2S}(x) = 2S\,H$ in (\ref{SZT0}) one immediately
finds\footnote{Note that these arguments are easily extended to the
case of non-zero temperatures.}
\begin{equation}
  v_{2S}\,\chi_{2S} + v_1\,\chi_{1} = {S\over\pi}\ .
\label{sumX}
\end{equation}
{}From the known susceptibilities for the limiting cases of the spin-$S$
and spin-$(S-1/2)$ Takhtajan-Babujian spin \cite{Babu83} chains we
expect that $(v_{2S}\chi_{2S})$ decreases continuously from $S/\pi$
for vanishing hole concentration $x_h$ to $(S-1/2)/\pi$ for $x_h=1$.
This implies $v_1\chi_1=0$ for $x_h=0$.  However, since $v_1$ vanishes
in this limit as well, these simple considerations do \emph{not} rule
out a finite or even singular contribution $\chi_1$ to the
susceptibility (\ref{susc0}).

%
\section{Calculation of the susceptibility}
\label{sec:RHP}
To calculate $\chi_j$ we have to analyze the Wiener-Hopf equations
(\ref{WHener}) for the dressed energies in the limit $H\to0$.  In
terms of the functions $f_j(x) = \epsilon_j(\Lambda_j+x)$ the WH
equations can be rewritten as ($t\equiv \Lambda_{2S}-\Lambda_1$)
\begin{eqnarray}
  f_1(x) &=& -2\pi A\,e^{-\pi \left|\Lambda_1+ x\right|}
	+ \int_{0}^\infty{\rm d}y\
	              K_{11}(x-y) f_1(y)
	+ \int_{0}^\infty{\rm d}y\
                      K_{12}(x-y-t) f_{2S}(y)\ ,
\nonumber\\
\label{WHener2}\\
  f_{2S}(x)  &=& {1\over2}H-2\pi\,e^{-\pi \left|\Lambda_{2S}+x\right|}
	+ \int_{0}^\infty{\rm d}y\
		    K_{21}(x-y+t) f_1(y)
	+ \int_{0}^\infty{\rm d}y\
	            K_{22}(x-y) f_{2S}(y)\ .
\nonumber
\end{eqnarray}
Again, the $\Lambda_j$ have to be determined such that $f_j(0)=0$.
By Fourier transformation of (\ref{WHener2}) we obtain
\begin{equation}
  G^T(\omega)\hat F_+(\omega)-\hat F_-(\omega)=T(\omega)\ .
\label{matrform}
\end{equation}
Here, $\hat{F}_\pm$ are the two-component vectors
\begin{equation}
  \hat{F}_\pm(\omega) = \pm\left(\begin{array}{l}
	\int_0^\infty {\rm d}x\ e^{\pm i\omega x} f_1(x)\\
	\int_0^\infty {\rm d}x\ e^{\pm i\omega x} f_2(x)
		\end{array}
        \right)
\end{equation}
of analytical functions of $\omega$ in the upper (lower) half-planes,
respectively.  $G(\omega)$ is a matrix given by
\begin{equation}
  G(\omega) =
   U^{-1}(\omega)\, \left( I - \hat{\mathbb{K}}^{(S)}(\omega)\right)\,
   U(\omega)\ ,
  \qquad
  \det\,G(\omega) = \left(1+e^{-|\omega|}\right)^{-2}
     \frac{\left(1-e^{-|\omega|}\right)}
           {\left(1-e^{-(2S-1)|\omega|}\right)}
\label{defG}
\end{equation}
(here $U(\omega) = {\rm diag} \left(\exp({-i\omega\Lambda_1}),
\exp({-i\omega\Lambda_{2S}})\right)$ and $I$ is the $2\times2$ unit
matrix) and $T(\omega)$ is the vector
\begin{equation}
  T(\omega) =
   \pi H \delta(\omega) \left(\begin{array}{c}0\\1\end{array}\right)
   -\frac{4\pi^2}{\omega^2+\pi^2}\ \Omega(\omega)\ ,\qquad
   \Omega(\omega) =
   U(\omega)\left(\begin{array}{c}A\\1\end{array}\right)\ .
\label{defT}
\end{equation}

Now the solution of the equation (\ref{matrform}) can be given in
terms of the one of the regular matrix Riemann--Hilbert problem (RHP):
\begin{eqnarray}
  &&Z(\omega)\to I,\qquad \omega\to\infty,
\nonumber\\
  &&Z(\omega)\quad\mbox{is analytical for}\quad \omega\notin\mathbb{R},
\label{origRHP}\\
  &&Z_-(\omega)=Z_+(\omega)G(\omega), \qquad \omega\in\mathbb{R}.
\nonumber
\end{eqnarray}
Here and below the subscripts $\pm$ denote the limit values of
functions  from the left (right) of the contour.

If the solution of the RHP (\ref{origRHP}) is found, then
Eq.~(\ref{matrform}) can be written in the form
\begin{equation}
\label{newform}
\Bigl[Z_+^T(\omega)\Bigr]^{-1}\hat F_+(\omega)-
\Bigl[Z_-^T(\omega)\Bigr]^{-1}\hat F_-(\omega)=
\Bigl[Z_-^T(\omega)\Bigr]^{-1}T(\omega),
\end{equation}
which is evidently solved by
\begin{equation}
\label{solution}
\hat F_+(\omega)=\frac{1}{2\pi i}
Z_+^T(\omega)\int_{-\infty}^\infty {\rm d}u\
\frac{\Bigl[Z_-^T(u)\Bigr]^{-1}T(u)}{u-\omega_+}\,.
\end{equation}
Now we can substitute here the explicit form (\ref{defT}) of
$T(\omega)$.  The integration of $\delta$-function is trivial.  For
the second term one can shift the integration contour in
(\ref{solution}) to the lower half-plane.  Then the only singularity
of the integrand is simple pole at the point $u=-i\pi$ and we obtain
\begin{equation}
\label{expsolution}
 \hat F_+(\omega)=\frac{iH}{2\omega_+} Z_+^T(\omega)
 \Bigl[Z_-^T(0)\Bigr]^{-1}\left(\begin{array}{c}0\\1\end{array}\right)\,
-
 \frac{2\pi i}{\omega+i\pi}\
 Z_+^T(\omega)\Bigl[Z_-^T(-i\pi)\Bigr]^{-1}\Omega(-i\pi)\ .
\end{equation}

The condition $f_j(x=0)=0$ implies $-i\lim_{\omega\to\infty}\omega
\hat F_+(\omega)=0$.  Hence the boundaries of integration $\Lambda_j$
are given as a funtion of the magnetic field $H$ and the parameter $A$
by
\begin{equation}
\label{zerocond}
  \Omega(-i\pi)=\frac{H}{4\pi}Z_-^T(-i\pi)
  \Bigl[Z_-^T(0)\Bigr]^{-1}
  \left(\begin{array}{c}0\\1\end{array}\right)\ .
\end{equation}

Similarly, the ground state energy (\ref{Emag}) is now
\begin{eqnarray}
  e_s &=&
   -2\pi\left(
    {e^{-\pi\Lambda_{2S}}\over v_{2S}}\int_0^\infty e^{-\pi x}f_{2S}(x)
   +{Ae^{-\pi\Lambda_1}\over v_1}\int_0^\infty e^{-\pi x}f_1(x)
  \right)
\nonumber\\[8pt]
  &=&  -2\pi \Omega(-i\pi)^T V^{-1} \hat F_+(i\pi)\ ,
\label{Emag2}
\end{eqnarray}
($V = {\rm
diag}\left(v_{1},v_{2S}\right)$ is a diagonal matrix containing the
Fermi velocities (\ref{veloS}) of the magnetic modes).
Using (\ref{zerocond}) in (\ref{expsolution}) we can express
$F_+(i\pi)$ through the solution of the RHP
\begin{equation}
\label{F+ipi}
  \hat F_+(i\pi)=\frac{H}{4\pi}Z_+^T(i\pi)\Bigl[Z_-^T(0)\Bigr]^{-1}
  \left(\begin{array}{c}0\\1\end{array}\right)\ .
\end{equation}
and finally have the magnetic susceptibility in terms of $Z$ as
\begin{equation}
  \chi = \frac{1}{4\pi} \left\{
    \Bigl[Z_-(0)\Bigr]^{-1}Z_-(-i\pi)\
    V^{-1}\ Z_+^T(i\pi)  \Bigl[Z_-^T(0)\Bigr]^{-1}
  \right\}_{22}\ .
\label{susc1}
\end{equation}

\begin{figure}[ht]
\begin{center}
\leavevmode
\epsfxsize=0.65\textwidth
(a)\epsfbox{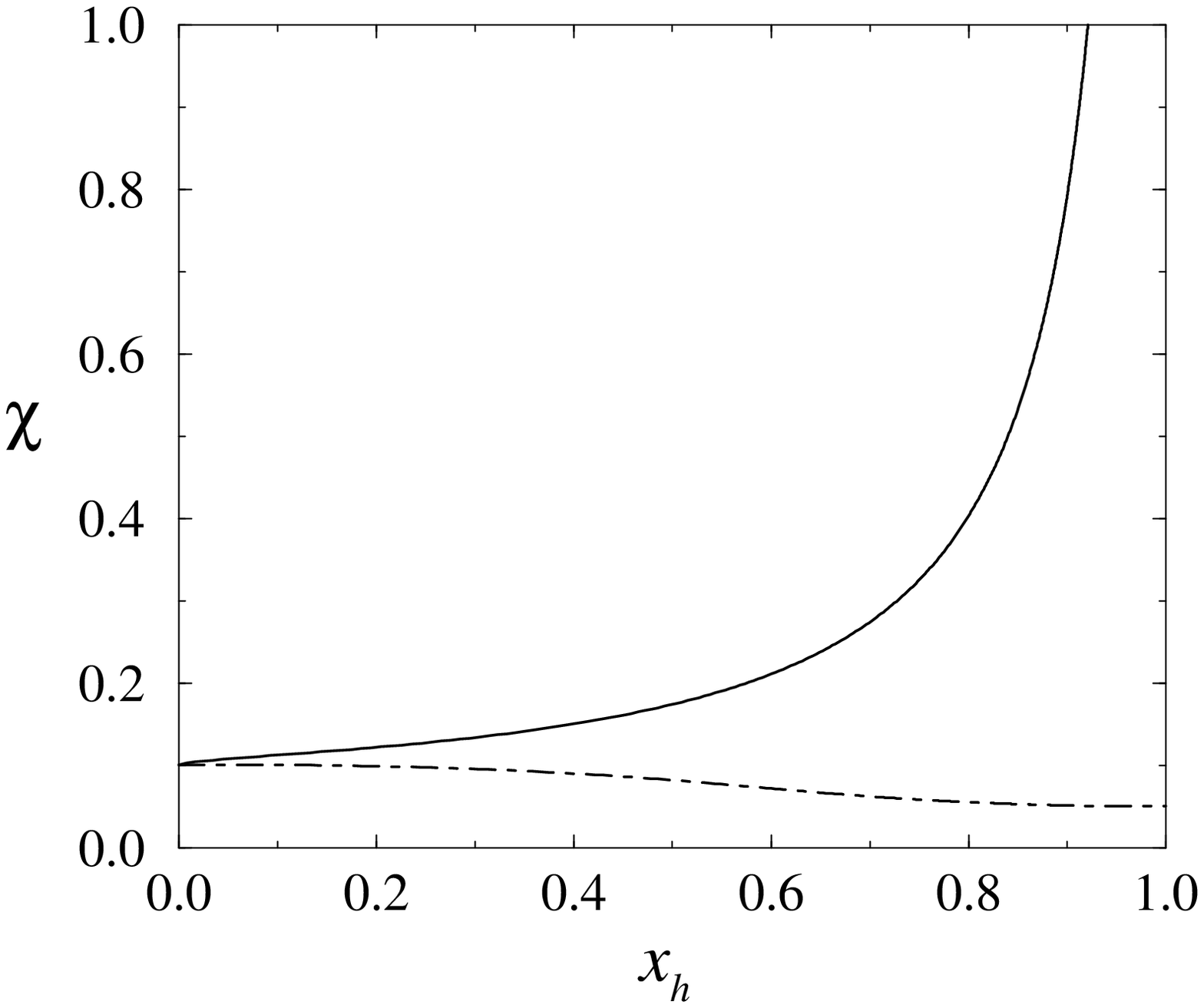}
\end{center}
\begin{center}
\leavevmode
\epsfxsize=0.65\textwidth
(b)\epsfbox{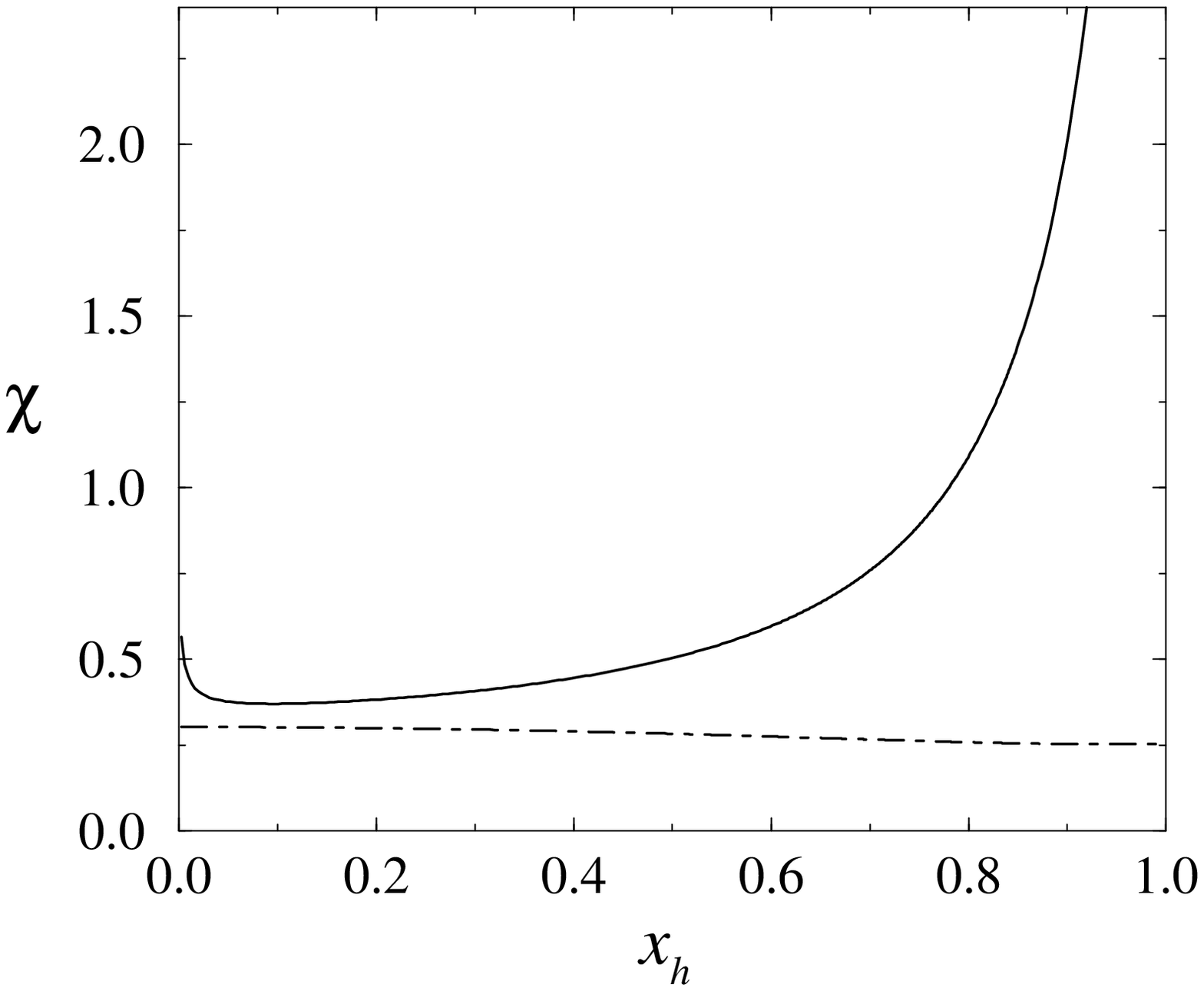}
\end{center}
\caption{Magnetic susceptibility of the doped $S=1$ (a) and $S=3$ (b)
system as a function of the concentration $x_h$ of holes (full line).
The broken line is the contribution of the `background' mode
$\epsilon_{2S}$ to $\chi$.
\label{fig:sus1x}}
\end{figure}

The symmetries the matrix $G(\omega)$ allow to deduce some general
properties of the solution $Z$ of the matrix Riemann--Hilbert problem
(\ref{origRHP}).  As a consequence of the identity $G^T(-\omega) =
G(\omega)$ one has $Z_+(\omega)\, Z_-^T(-\omega) = I$.
This allows to rewrite the expression (\ref{susc1}) in terms of
$Z_+(0)$ and $Z_+(i\pi)$ alone
\begin{equation}
\label{susc2}
\chi= \frac{1}{4\pi} \left\{G^{-1}(0)
  \Bigl[Z_+(0)\Bigr]^{-1}Z_+^T(i\pi)V^{-1} \left(
  \Bigl[Z_+(0)\Bigr]^{-1}Z_+^T(i\pi)\right)^{-1}
  \right\}_{22}\ .
\end{equation}
Eqs.~(\ref{defG}) and (\ref{kernS}) yield
\begin{equation}
\label{G0s}
G(0)=\frac{1}{2(2s-1)}\left(
  \begin{array}{cc} 2s& -1\\ -1& 1 \end{array} \right),\qquad
G^{-1}(0)=2\left(
  \begin{array}{cc}1& 1\\ 1& 2s \end{array} \right)\ .
\end{equation}
Replacing $V$ by a unit matrix this immediately reproduces the
susceptibility sum rule (\ref{sumX}).  In Figure~\ref{fig:sus1x} we
present numerical results from the integration of Eqs.~(\ref{WHener2})
for $S=1$ and $S=3$, respectively.  In the limiting cases $x_h\to0$
and $x_h\to1$ corresponding to undoped spin-$S$ and -$(S-1/2)$ chains,
the velocity $v_1$ vanishes.  The singular behaviour of the
susceptibility near $x_h=1$ is a consequence of the contribution
$\chi_1\sim 1/(2\pi v_1)$ to the susceptibility.

\section{Susceptibility at low doping}
\label{sec:smallx}
Interestingly, however, the dependence of the susceptibility on the
hole concentration is found to depend strongly on the value of $S$ for
$x_h\to0$: for all values of $S$ the velocity $v_1$ vanishes with
$x_h$.  In fact, since $x_h\ll1$ corresponds to small $Q$ in the
integral equations (\ref{rho0}) and (\ref{kappa0}) for the densities
and dressed energies of the charge excitations, the latter can be
solved by iteration giving ($\psi(x)$ and $\psi^{(2)}(x)$ are the
digamma and polygamma functions, respectively)
\begin{equation}
  A = {\pi^2\over12}
      { \psi^{(2)}\left({2S+3\over4}\right) -
	\psi^{(2)}\left({2S+1\over4}\right) \over
	\psi\left({2S+3\over4}\right) -
	\psi\left({2S+1\over4}\right) }\,\,x_h^3
\end{equation}
and $v_1=\pi A/x_h \propto x_h^2$.  As discussed above, we expect
$v_1\chi_1\to0$ in the limit $x_h\to0$ which agrees with numerical
findings.  The contribution $\chi_1$ of the `minimal' mode
$\epsilon_1(x)$ alone to the susceptibility, however, is singular for
sufficiently large value of $S$.

In general, however, we have $v_1\ne v_{2S}$ and to find the
susceptibility we need to solve matrix RHP~(\ref{origRHP}).  As in the
studies of the low temperature thermodynamics \cite{FrPT98,Frahm99} we
make use of the fact that we have $A\ll1$ for small hole concentration
$x_h$: for small $H$ one finds one has $0\ll\Lambda_1\ll\Lambda_{2S}$
in this regime.  Thus, the jump matrix $G(\omega)$~(\ref{defG})
contains the \emph{large} parameter $t\to+\infty$.  This allows to
find the solution of the RHP~(\ref{origRHP}) asymptotically and thus
to determine the susceptibility as a function of $A$.  Since the
treatment for $S=1$ and $S>1$ differs in we present the main results
of this asymptotic analysis separately below.  Further technical
details are presented in the Appendix.

\subsection{$S=1$}
To solve the RHP~(\ref{origRHP}) we factorize its solution into a
product of two matrices
\begin{equation}
\label{subst}
   Z(\omega)=\Phi(\omega)\left(
	\begin{array}{cc}
	1&0\\0&\alpha(\omega)
	\end{array}\right).
\end{equation}
Here the function $\alpha(\omega)$ is analytical for $\omega\notin
\mathbb{R}$, $\alpha(\omega)\to1$ as $\omega\to\infty$, and solves the
regular scalar RHP on the real axis:
\begin{eqnarray}
  \alpha_-(\omega)&=&\alpha_+(\omega)\,\det\,G(\omega)
\nonumber\\
  &=&\alpha_+(\omega) \left(1+e^{-|\omega|}\right)^{-2} \mbox{~for~}
  \omega\in \mathbb{R},
\label{rhpa}
\end{eqnarray}
It is easy to see that
\begin{equation}
\label{solaexp}
  \alpha_+(\omega)=\frac{2\pi}{\Gamma^2\left(
  \frac{1}{2}-\frac{i\omega}{2\pi}\right)}
  \left(-\frac{i\omega}{2\pi e}\right)^{-\frac{i\omega}{\pi}},
\end{equation}
and $\alpha_-(\omega)= \alpha_+^{-1}(-\omega)$.  In particular, we have
$\alpha_+(0)=2$ and $\alpha_+(i\pi)={\pi}/{e}$.

{}From (\ref{origRHP}) we find that the matrix $\Phi(\omega)$
in~(\ref{subst}) is analytical for $\omega\notin\mathbb{R}$, it
approaches identity at $\omega\to\infty$ and its boundary values on
the real axis are related by
\begin{equation}
  \Phi_-(\omega)=\Phi_+(\omega)G_\Phi(\omega),
  \qquad \omega\in \mathbb{R}\ .
\label{RHPPhi}
\end{equation}
The matrix
\begin{equation}\label{GPhi}
 G_\Phi(\omega)=\left(
 \begin{array}{cr}
\displaystyle 1 &\quad\displaystyle -\frac{\alpha_-^{-1}(\omega)}
{2\cosh\frac{\omega}{2}} e^{-i\omega t}\\
\displaystyle
\displaystyle -\frac{\alpha_+(\omega)}
{2\cosh\frac{\omega}{2}} e^{i\omega t}&\quad
1+e^{-|\omega|}
\end{array}
\right)
\end{equation}
can be factorized into the product $G_\Phi(\omega)= M_+(\omega)
M_-(\omega)$ with ($\sigma_\pm$ are Pauli matrices)
\begin{equation}
\label{M+-}
  M_+(\omega)=I-\frac{\alpha_+(\omega)}
  {2\cosh\frac{\omega}{2}} e^{i\omega t}\sigma_-,\qquad
  M_-(\omega)=I-\frac{\alpha_-^{-1}(\omega)}
  {2\cosh\frac{\omega}{2}} e^{-i\omega t}\sigma_+\ .
\end{equation}
Obviously $M_\pm(\pm\omega)$ are analytical in the strip
$0<\Im\omega<\pi$.  In particular $M_\pm(\omega)$ can be analytically
continued into the strip $0<\pm\Im\omega<a$ for some $a\in(0,\pi)$
where the off-diagonal entries of $M_\pm$ vanish in the limit
$t\to\infty$.  Following the asymptotic analysis of a similar problem
in Ref.~\cite{DZ} we introduce a deformation of the RHP
(\ref{RHPPhi}) which can be solved in terms of a absolutely convergent
series in $\exp{(-\pi t)}$.  In Appendix~\ref{app:S1} the asymptotic
behaviour of the matrix $\Phi(\omega)$ in the points $\omega=i\pi,~0$
is computed from this deformed RHP ($p$ and $q$ are functions of $t$
defined in Eq.~(\ref{pq}))
\begin{equation}
\label{Phi+ipi}
\Phi_+(i\pi)=\left(
\begin{array}{cc}
  \displaystyle 1+p^2(q+1) &\displaystyle p\\
  \displaystyle qp & \displaystyle 1-p^2
\end{array}\right)+o\left(p^2\right),
\end{equation}
\begin{equation}
\label{estPhi0}
\Phi_+(0)=\left(
\begin{array}{cc}
  \displaystyle 1+2p-2p^2 &\displaystyle 2p\\
  \displaystyle 1-2p-2p^2 & \displaystyle 1-2p^2
\end{array}\right)+o\left(p^2\right).
\end{equation}
Using Eqs.~(\ref{zerocond}), (\ref{susc2}) we obtain $A=2\pi p^2
+o\left(p^2\right)$ which finally gives the zero field limit of the
magnetic susceptibility for small doping (or, equivalently, $A\ll1$)
\begin{equation}
\label{susc45}
  \chi= \frac{1}{\pi}
	\left(\frac{1-\xi}{v_2} + \frac{\xi}{v_1}\right),
\end{equation}
where
\begin{equation}
\label{xi}
  \xi=-\frac{A}{2\pi}\left(\log \frac{A}{2\pi} +2\mathbf{C}\right)
      +o(A)\ .
\end{equation}
This result refines the estimate given in Ref.~\cite{FrPT98}.  In
terms of the hole concentration the contribution $\chi_1$ to the
magnetic susceptibility is
\begin{equation}
\label{chi1}
  \chi_1 = \frac{3x_h}{2\pi^3}\left(\log x_h + 0.217927\ldots\right)
\end{equation}
which is in excellent agreement with the numerical data in
Figure~\ref{fig:sus1x}(a).
%
\subsection{$S>1$}
Similar to the case $S=1$ the solution of the corresponding RHP is
given by the product of two matrices
\begin{equation}
\label{substs}
  Z(\omega)=\Phi(\omega)\left(
	\begin{array}{cc}
	\beta_1(\omega)&0\\
	0&\beta_2(\omega)
	\end{array}\right),
\end{equation}
where $\beta_j(\omega)$ solve scalar regular RHPs with canonical
normalization condition and
\begin{eqnarray}
\beta_{1-}(\omega)&=&\beta_{1+}(\omega)G_{11}(\omega),
\qquad \omega\in \mathbb{R},
\nonumber\\
\beta_{2-}(\omega)&=&\beta_{2+}(\omega)
\frac{\det G(\omega)}{G_{11}(\omega)},
\qquad \omega\in \mathbb{R}.
\label{RHPb}
\end{eqnarray}
Below, we will need only the ratio $\beta_2(\omega)/\beta_1(\omega)
\equiv \alpha^{(S)}(\omega)$, which is equal to
\begin{equation}\label{explas}
  \alpha^{(S)}_+(\omega)=
  \frac{\displaystyle(2S-1)^{\frac{i\omega}{2\pi}(2S-1)+\frac12}}
       {\displaystyle(2S)^{\frac{2i\omega S}{\pi}+1}}
  \cdot
  \frac{\Gamma\left(-\frac{i\omega}{2\pi}\right)
  \Gamma\left(-\frac{i\omega}{2\pi}(2S-1)\right)}
  {\Gamma^2\left(-\frac{i\omega S}{\pi}\right)}
  \left(-\frac{i\omega}{2\pi e}\right)^
  {-\frac{i\omega S}{\pi }}.
\end{equation}
As before $\alpha^{(S)}_+(\omega)\alpha^{(S)}_-(-\omega)=1$ and in
particular
\begin{equation}\label{partals}
\alpha^{(S)}_+(0)=2S\eta,\qquad
\alpha^{(S)}_+\left(\frac{i\pi}{S}\right)=
\frac{\pi}{e\sin\frac{\pi}{2S}}\eta^{\frac{S-1}{S}},
\qquad \eta=(2S-1)^{-1/2}\ .
\end{equation}

The asymptotic analysis of the RHP for $\Phi(\omega)$ is again based
on the factorization of the corresponding jump-matrix $G_\Phi(\omega)$
and subsequent deformation of the original jump-contour (see
Appendix~\ref{app:Sx} for details).  The matrices $\Phi_+(0)$ and
$\Phi_+(i\pi)$ needed for the calculation of the susceptibility are
\begin{equation}\label{Phi0s}
\Phi_+(0)=\left(
\begin{array}{ccc}
 1+u_1p\eta-u_1^2p^2/2&\quad&
 u_1p\\
(1-u_1^2p^2/2)\eta-u_1p&\quad&
 1-u_1^2p^2/2
\end{array}\right)+o(p^2),
\end{equation}
and
\begin{equation}\label{Phi+ipis}
\Phi_+(i\pi)=\left(
\begin{array}{lr}
 1+\frac{u_1^2p^2}{2(S-1)}
&
 \frac{u_1p}{S+1}-\frac{u_2p^2}{S+2}\\
 \frac{u_1p}{S-1}-\frac{u_2p^2}{S-2}
&
 1-\frac{u_1^2p^2}{2(S+1)}
\end{array}
\right)+o(p^2).
\end{equation}
Here we have used the following notations
\begin{equation}\label{ugamma}
  u_k=\frac{1}{\pi}\,\alpha^{(S)}_+\left(\frac{i\pi k}{S}\right)
  \sin\left(\frac{\pi k}{2S}\right),
  \qquad p=e^{-\frac{\pi t}{S}}.
\end{equation}
In the case $S=2$ the entry $\Phi_{21+}(i\pi)$ should be replaced
by
\begin{equation}
\label{result21}
  \Phi_{21+}(i\pi)=u_1p-\pi u_2p^2\left(t-
  i\left.\partial_z \log\alpha^{(S)}_+(z)\right|_{z=i\pi} \right)
  +o(p^2),\qquad S=2.
\end{equation}
This difference, however, does not affect the final results for $A$
which is
\begin{equation}
\label{Atts}
  A=\frac{S\alpha^{(S)}_+(i\pi)}{(S+1)e}
  \eta^{\frac{S-1}{S}}p^{S+1}+o(p^{S+1})\ .
\end{equation}
%
%
%
For susceptibility of the spin-$S$ model at small hole concentration
we obtain
\begin{equation}
\label{susc44s}
\chi = \frac{S}{\pi}\left(
\frac{1-\xi}{v_{2S}}+\frac{\xi}{v_{1}}\right),
\end{equation}
with
\begin{equation}
\label{xiss}
\xi=\frac{S^2}{S^2-1}(u_1p)^2  +o(p^2).
\end{equation}
Substituting here $u_1$ from~(\ref{ugamma})
and $p$ from~(\ref{Atts}) we finally obtain
\begin{equation}
\label{xifinals}
  \xi=\frac{1}{S-1}
  \left(\frac{\Gamma^2(S)}{\Gamma(1/2)\Gamma(S-1/2)}
  \right)^{\frac{2}{S+1}}
  \left(\frac{2S-1}{4S^2(S+1)}\right)^{\frac{S-1}{S+1}}\,
  A^{\frac{2}{S+1}} + o(A^{\frac{2}{S+1}})\ .
\end{equation}
%
%
%
As a function of the hole concentration the contribution of the
minimal mode $\epsilon_1$ to the susceptibility is $\chi_1\propto
x_h^{2\frac{2-S}{S+1}}$, which will give a singularity for $S>2$.  For
$S=3$, the case depicted in Fig.~\ref{fig:sus1x}(b), $\chi_1$ diverges
as $1/\sqrt{x_h}$ for small $x_h$.

\section{Summary and Conclusions}
\begin{figure}[ht]
\begin{center}
\leavevmode
\epsfxsize=0.65\textwidth
\epsfbox{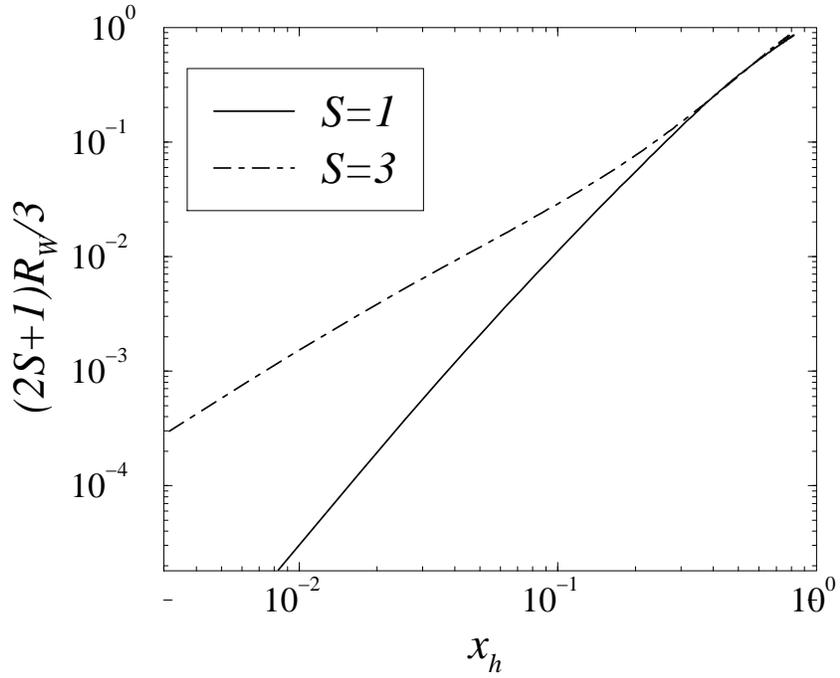}
\end{center}
\caption{Wilson ratio (\protect{\ref{wilson}}) as a function of
the hole concentration $x_h$ for the $S=1$ and $S=3$
system.
\label{fig:wilson}}
\end{figure}

In this paper we have studied the magnetic properties of a class of
integrable models for doped spin-$S$ Heisenberg chains in the limit of
very small magnetic field.  Of particular interest was the singular
behaviour of the susceptibility as a function of the hole
concentration for small $x_h$.  The response of the two magnetic modes
present in these systems to an external field is essential for the
proper identification of an effective field theoretical description of
the system which in turn will allow for analytical investigations of
the properties of the metallic phases of quasi-one dimensional
transition metal oxides.  The susceptibility due to the `background
mode' $\epsilon_{2S}(x)$, which interpolates smoothly between the
spinons of the spin-$S$ and -$(S-1/2)$ Takhtajan-Babujian models when
the hole concentration is varied between $x_h=0$ and $x_h=1$,
decreases monotonically from $S/\pi$ to $(S-1/2)/\pi$ as a function of
$x_h$.
The contribution $\chi_1$ of the second magnetic mode $\epsilon_1(x)$
can be singular near $x_h=0,1$ as a consequence of the vanishing of
the corresponding Fermi velocity.  For $x_h\to1$, where few spin-$S$
particles propagate in a background of the spin-$(S-1/2)$ holes this
singularity may be understood in a similar way as for an underscreened
Kondo impurity (although the net impurity moment vanishes in the
ground state).  For small doping the effect of the doping on the
susceptibility is more subtle.  Our analysis shows that
$\lim_{x_h\to0}\chi_1$ is non-zero only for $S\ge2$.  The effect of
the (non-universal) velocities on the singular behaviour of the
susceptibility (\ref{susc0}) and the specific heat (\ref{spech}) can
be eliminated by introducing a ``Wilson Ratio'' for the relative
contributions of the two magnetic modes
\begin{equation}
\label{wilson}
  R_W = \frac{\chi_1/\chi_{2S}}{C_{1}/C_{2S}}\ .
\end{equation}
With this definition the contribution from $\epsilon_{2S}$ is
interpreted as that of the host while associating the mode
$\epsilon_1$ with the impurities which is certainly justified in the
limits $x_h\to 0,1$.  In these limits we find
\begin{equation}
  R_W \simeq \left\{
   \begin{array}{ll}
   \frac{3S(2S+1)}{(S+2)(2S-1)}\,\xi & x_h\to 0\\
   \frac{3}{2S+1} & x_h\to 1
   \end{array}
   \right. ,
\end{equation}
where according to (\ref{xi}) and (\ref{xifinals}) $\xi$ vanishes as
$x_h^3\log x_h$ for $S=1$ and as $x_h^{\frac{6}{S+1}}$ for $S>1$.
Extending the interpretation of the two magnetic modes to arbitrary
hole concentration one finds a continuous change of $R_W$ with $x_h$
(see Figure~\ref{fig:wilson}): as was to be expected from the low
temperature specific heat (\ref{spech}) the ratio defined in
(\ref{wilson}) is not universal.

Note that our results are the leading contributions to the
susceptibility for small magnetic field.  Due to approximations such
as that of Eqs.~(\ref{intds}) by (\ref{WHdens}) additional terms
$\propto 1/\log H$ to the susceptibility have been neglected.  For the
undoped systems the precise form of these contributions (which should
coincide with the ones present in $\lim_{x_h\to0}\chi_{2S}$) is given
in Ref.~\cite{Babu83}.

Finally, we mention that the solution of the matrix Riemann-Hilbert
problem (\ref{origRHP}) presented in this paper opens the possibility
to study the critical exponents in the asymptotics of correlation
functions in the $H=0$ phase of the doped Heisenberg chains.  In the
case of a simple Tomonaga-Luttinger liquid the spin part of
correlation functions is completely determined by the $SU(2)$ symmetry
and only the compactification radius of the boson related to the
charge mode gives rise to anomalous exponents \cite{FrKo90}.  In the
doped spin-$S$ chains considered here one should expect an additional
doping dependence of the critical exponents due to the presence of two
magnetic modes: in spite of the $SU(2)$ invariance of the system a
subtle balance between these modes is preserved which is manifest in
the sum rules for the low temperature specific heat (\ref{sumC}) and
the zero field magnetic susceptibility (\ref{sumX}).  Further
investigations are also necessary to eludicate the relation of these
quantities in the finite field phase and the one with vanishing
magnetic field.

\section*{Acknowledgements}
We would like to thank A.\,R.~Its for useful discussions.  This work
has been supported by the Deutsche Forschungsgemeinschaft under Grant
No.\ Fr~737/2.

\newpage
\appendix
\section{Asymptotic analysis of the RHP for large $t$}
\subsection{$S=1$}
\label{app:S1}
\begin{figure}[ht]
\begin{center}
\begin{picture}(370,160)
\put(0,80){\vector(1,0){335}}
\put(150,0){\vector(0,1){200}}
\put(290,84){$\mathbb{R}$}
{\thicklines
\put(10,120){\line(1,0){335}}
\put(10,40){\line(1,0){335}}
\put(190,120){\vector(1,0){10}}
\put(200,40){\vector(-1,0){10}}
\put(148,118.5){$\scriptscriptstyle\bullet$}
\put(148,139){$\scriptscriptstyle\bullet$}
\put(148,38.5){$\scriptscriptstyle\bullet$}
\put(148,19){$\scriptscriptstyle\bullet$}
\put(155,123){$ia$}
\put(155,143){$i\pi$}
\put(126,43){$-ia$}
\put(126,23){$-i\pi$}
}
\put(290,124){$\Gamma_+$}
\put(290,44){$\Gamma_-$}
\put(180,100){$U=\Phi M_+$}
\put(180,53){$U=\Phi M_-^{-1}$}
\put(180,150){$U=\Phi$}
\put(180,0){$ U=\Phi$}
\end{picture}
\end{center}
\caption{The jump-contour for the new RHP (\protect{\ref{rhpU}}).
\label{contour}}
\end{figure}
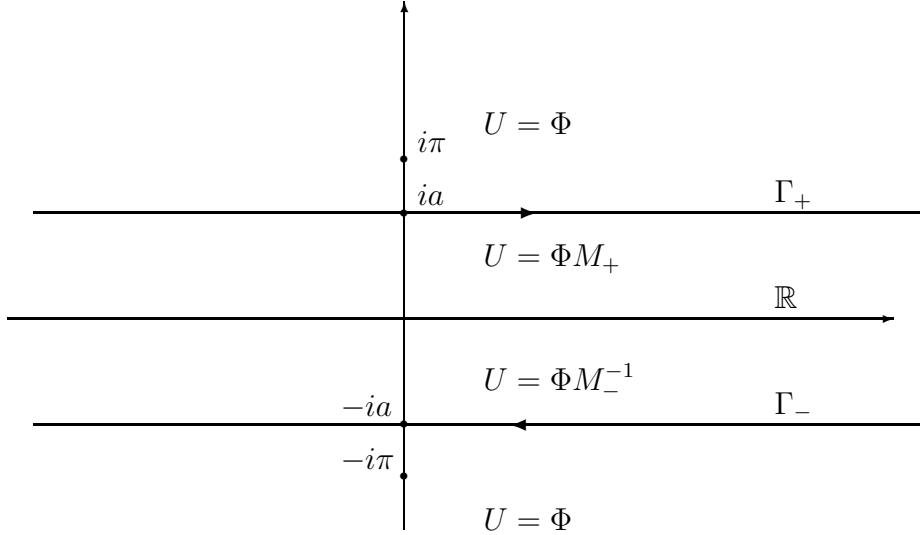

Using the factorization of the jump matrix $G_\Phi(\omega)$ in
Eq.~(\ref{RHPPhi}) we follow Ref.~\cite{DZ} by defining a new
matrix $U(\omega)$ as
\begin{equation}\label{defU}
U(\omega)=\left\{
  \begin{array}{lr}
    \Phi(\omega), & |\Im\omega|>a,\\
    \Phi(\omega)M_+(\omega),& 0<\Im\omega<a,\\
    \Phi(\omega)M_-^{-1}(\omega),& -a<\Im\omega<0,
\end{array}\right.
\end{equation}
where $a\in (0,\pi)$ (see Fig.~\ref{contour}).
Then $U(\omega)$ solves the following regular RHP:
\begin{equation}\label{rhpU}
\begin{array}{l}
\displaystyle
U(\omega)\to I,\qquad \omega\to\infty,\\
\displaystyle
U(\omega)\quad\mbox{is analytical for}\quad \omega\notin \Gamma,\\
\displaystyle
U_-(\omega)=U_+(\omega)G_U(\omega), \qquad
\omega\in \Gamma,
\end{array}
\end{equation}
where we have introduced
\begin{equation}\label{GU}
\displaystyle
G_U(\omega)=\left\{
  \begin{array}{lr}
  M_+(\omega), \qquad &\omega\in\Gamma_+\\
  M_-^{-1}(\omega), \qquad &\omega\in\Gamma_-
\end{array}
\right.\ .
\end{equation}
The contour $\Gamma=\Gamma_+\cup\Gamma_-$ is shown in
Fig.~\ref{contour} (the arrows show positive direction).  It is easy
to see that $U(\omega)$ has no cut on the real axis.  Thus
$U_-(\omega)$ can be analytically continued inside the strip
$|\Im\omega|<a$.  Similarly, $U_+(\omega)$ is analytical in the
remaining domain $|\Im\omega|>a$, where it coincides with the matrix
$\Phi(\omega)$ (more precisely $U_+(\omega)=\Phi_+(\omega)$ for
$\Im\omega>a$, but $U_+(\omega)=\Phi_-(\omega)$ for $\Im\omega<-a$).

{}From the explicit expression (\ref{M+-}) we have $M_\pm-I\sim e^{-at}$
on the contour $\Gamma_\pm$.  This implies
\begin{equation}\label{solU}
U(\omega)=I+o\left(e^{-at}\right),\qquad t\to\infty,
\end{equation}
for the solution of (\ref{rhpU}) where $a$ is arbitrary from the
interval $(0,\pi)$.  Thus, for $t\to\infty$ the asymptotic behaviour
of the solution to (\ref{RHPPhi}) is
\begin{equation}\label{solPhi}
\begin{array}{l}
\Phi(\omega)=I+O\left(e^{-\pi t}\right),\qquad |\Im\omega|>\pi,\\
\Phi(\omega)=I+O\left(e^{-\omega_0t}\right),
\qquad 0<|\Im\omega|=\omega_0\le\pi.
\end{array}
\end{equation}

This accuracy, however, is not sufficient to calculate the leading
doping dependence of the susceptibility.  In particular this estimate
can not be used to determine $\Phi(0)$ entering Eqs.~(\ref{zerocond})
and (\ref{susc2}).  To obtain the subleading contributions we can use
singular integral equations equivalent to the RHP ~(\ref{rhpU}).  Note
that only $\Phi_+(i\pi)=U_+(i\pi)$ and $\Phi_+(0)$ are needed for the
susceptibility.

{}From (\ref{GPhi}) we find $\det G_\Phi(\omega)=1$ and
$G_\Phi(\omega)=G_\Phi^T(-\omega)$, hence we conclude that
\begin{equation}
\label{condPhi}
  \det\Phi(\omega)=1\ , \qquad
  \Phi_+(\omega)\Phi_-^T(-\omega)=I\ .
\end{equation}
Hence we have
\begin{equation}\label{Phiipi}
\begin{array}{ll}
\Phi_{11+}(i\pi)=U_{11+}(i\pi)=U_{22+}(-i\pi),&\qquad
\Phi_{21+}(i\pi)=U_{21+}(i\pi)=-U_{12+}(-i\pi),\\
\Phi_{12+}(i\pi)=U_{12+}(i\pi)=-U_{21+}(-i\pi),&\qquad
\Phi_{22+}(i\pi)=U_{22+}(i\pi)=U_{11+}(-i\pi)
\end{array}
\end{equation}
at $\omega=i\pi$ while $\Phi(\omega=0)$ is parameterized by a single
parameter $\varphi$
\begin{equation}\label{solPhi0}
\Phi_+(0)=\left(
\begin{array}{cc}
\cos\varphi+\sin\varphi&
\sin\varphi\\
\cos\varphi-\sin\varphi&
\cos\varphi
\end{array}\right)
\end{equation}
as a consequence of Eq.~(\ref{condPhi}) together with jump
condition~(\ref{RHPPhi}).

Consider now the singular integral equation for $U_+(\omega)$:
\begin{equation}\label{intU1}
U_+(\omega)=I-\frac{1}{2\pi i}\int_\Gamma
\frac{U_+(z)(G_U(z)-I)}{z-\omega_+}\,dz,
\end{equation}
where $\omega_+$ means that $\omega$ is shifted from the integration
contour to the left.  In components, the equations for the entries
$U_{21+}(\omega)$ and $U_{22+}(\omega)$ read:
\begin{eqnarray}
U_{21+}(\omega)&=&
\frac{1}{2\pi i}\int_{\Gamma_+}
\frac{U_{22+}(z)\alpha_+(z)e^{izt}}
{2\cosh\frac{z}{2}}\,\frac{dz}{z-\omega_+},
\nonumber\\
\label{intU21}
\\
U_{22+}(\omega)&=&
1-\frac{1}{2\pi i}\int_{\Gamma_-}
\frac{U_{21+}(z)e^{-izt}}
{2\alpha_-(z)\cosh\frac{z}{2}}\,\frac{dz}{z-\omega_+}.
\nonumber
\end{eqnarray}
Let $\omega\in \Gamma_-$ in the first of~(\ref{intU21}).  Then
shifting the integration contour into the upper half-plane, we obtain
\begin{equation}\label{sum2U21}
U_{21+}(\omega)=i\sum_{k=0}^{\infty}(-1)^k
\frac{U_{22+}\Bigl(i\pi(2k+1)\Bigr)
\alpha_+\Bigl(i\pi(2k+1)\Bigr)e^{-t\pi(2k+1)}}
{\omega-i\pi(2k+1)}.
\end{equation}
Since $U_{22+}(\omega)\to1$ and $\alpha_{+}(\omega)\to1$ at
$\omega\to\infty$, the series~(\ref{sum2U21}) is absolutely convergent
in the domain $\Im\omega<\pi$, and we arrive at
\begin{equation}\label{estU21}
U_{21+}(\omega)=
i\frac{U_{22+}(i\pi)\alpha_+(i\pi)e^{-t\pi}}
{\omega-i\pi}+O\left(e^{-3\pi t}\right),
\qquad \Im\omega<\pi.
\end{equation}

Substituting~(\ref{estU21}) into the equation for
$U_{22+}$, putting $\omega=-i\pi$ and shifting the integration contour
to the lower half-plane, we obtain
\begin{equation}\label{estU22+}
U_{22+}(-i\pi)=U_{11+}(i\pi)=
1+\frac{\pi}{2}U_{22+}(i\pi)e^{-2\pi t-2}
\left(t+\gamma+\frac{1}{2\pi}\right)+o\left(e^{-2\pi t}\right),
\end{equation}
where ($\mathbf{C}\approx0.57721\ldots$ is Euler's constant)
\begin{equation}\label{gamma}
  \gamma=-i\left.\partial_z\log\alpha_+(z)\right|_{z=i\pi}
  =\frac{1}{\pi}\left(\log2-\mathbf{C}\right)\ .
\end{equation}
Similarly, we  analyse the equations for $U_{11+}$ and $U_{12+}$ and
obtain
\begin{eqnarray}
U_{11+}(-i\pi)=U_{22+}(i\pi)&=&
 1-\frac{1}{2}U_{12+}(i\pi)e^{-\pi t-1}+o\left(e^{-2\pi t}\right),
\nonumber\\
U_{12+}(-i\pi)=-U_{21+}(i\pi)&=&
 -\pi e^{-\pi t-1}(t+\gamma)+o\left(e^{-2\pi t}\right).
\label{estU11}
\end{eqnarray}
Finally, using relations (\ref{Phiipi}) with Eqs.~(\ref{estU21}),
(\ref{estU22+}) and (\ref{estU11}) we find
\begin{equation}
\label{U+ipi}
  U_+(i\pi)=\Phi_+(i\pi)=\left(
  \begin{array}{cc}
	1+p^2(q+1) & p\\
	qp & 1-p^2
  \end{array}\right)+o\left(p^2\right),
\end{equation}
where $p$ and $q$ are defined as
\begin{equation}
\label{pq}
  p=\frac{1}{2}\,e^{-\pi t-1}\ ,\qquad
  q=2\pi t+2(\log2-\mathbf{C})\ .
\end{equation}

Since the estimate~(\ref{estU21}) is uniform in the domain
$\Im\omega<\pi$, we have with (\ref{defU})
\begin{equation}
\label{estphi210}
	\Phi_{21-}(0)=U_{21+}(0)=
	-U_{22+}(i\pi)e^{-t\pi-1}
	+{\cal O}\left(e^{-3\pi t}\right).
\end{equation}
Comparing with~(\ref{solPhi0}) and using~(\ref{U+ipi}) we find
$\sin\varphi=2p$, and thus we arrive at~(\ref{estPhi0}).

\subsection{$S>1$}
\label{app:Sx}
The asymptotic analysis of the RHP~(\ref{origRHP}) for $S>1$ is quite
similar to the case $S=1$.  Now the boundary values of the matrix
$\Phi(\omega)$ introduced in (\ref{substs}) on the real axis satisfy
\begin{equation}\label{RHPPhis}
\Phi_-(\omega)=\Phi_+(\omega)G_\Phi(\omega),
\qquad \omega\in \mathbb{R}
\end{equation}
with
\begin{equation}\label{Gphisexp}
G_\Phi(\omega)=\left(
\begin{array}{cc}
1&\qquad -\frac{1}{\alpha^{(S)}_-(\omega)}
\frac{\sinh\frac{\omega}{2}}{\sinh\omega S}
e^{-i\omega t}\\
-\alpha^{(S)}_+(\omega)
\frac{\sinh\frac{\omega}{2}}{\sinh\omega S}
e^{i\omega t}&\qquad\
e^{-\frac{|\omega|}{2}}\frac{\sinh\omega S}
{\sinh\omega(S-1/2)}
\end{array}
\right).
\end{equation}
As for $S=1$ this matrix can be factorized into the product of
matrices $M_+$ and $M_-$, which now are
\begin{equation}\label{Mspl}
M_+(\omega)=I-\alpha^{(S)}_+(\omega)
\frac{\sinh\frac{\omega}{2}}{\sinh\omega S}
e^{i\omega t}\sigma_-,\qquad
M_-(\omega)=I-\frac{1}{\alpha^{(S)}_-(\omega)}
\frac{\sinh\frac{\omega}{2}}{\sinh\omega S}
e^{-i\omega t}\sigma_+
\end{equation}
%

The following considerations almost completely repeat the
corresponding part of the subsection $S=1$.  First, we again come to
the new RHP on the contour $\Gamma$.  The only difference is that now
$a\in(0,\frac\pi{S})$.  Thus we obtain that $\Phi(\omega) \approx I$
up to exponentially small corrections everywhere exept
$\omega\in\mathbb{R}$.  In order to improve this estimate we use the
corresponding singular integral equations.  Now the asymptotic
behaviour of $\Phi(\omega)$ is defined by residues of the integrands
in the points $\omega=\frac{i\pi}S k$. Respectively the solution of
the RHP is given by the asymptotic series of $\exp\{-\frac{\pi t}S
k\}$.  Note also that just like in the case $S=1$ the matrix $\Phi(0)$
can be parametrized by a single parameter.  The explicit expression
now has the form
\begin{equation}\label{solPhi0s}
\Phi_+(0)=\left(
\begin{array}{cc}
\cos\varphi+\eta\sin\varphi&
\quad\sin\varphi\\
\eta\cos\varphi-\sin\varphi&
\quad\cos\varphi
\end{array}\right).
\end{equation}

\setlength{\baselineskip}{14pt}

%


%
%

\end {document}